\documentclass{aa}  

\usepackage{graphicx}
\usepackage{txfonts}
\usepackage{comment}
\usepackage{hyperref}
\usepackage{amsmath}
\usepackage{float}
\newcommand{\RomanNumeralCaps}[1]
    {\MakeUppercase{\romannumeral #1}}
\begin{document}

   \title{Deep learning image burst stacking to reconstruct high-resolution ground-based solar observations}


   \author{C.Schirninger
          \inst{1}
          \and
          R. Jarolim\inst{2}
          \and
          A. M. Veronig\inst{1,}\inst{3}
          \and
          C. Kuckein\inst{4,}\inst{5,}\inst{6}
          }

   \institute{Institute of Physics, University of Graz,
              Universitätsplatz 5, 8010 Graz, Austria
         \and
             High Altitude Observatory, National Center for Atmospheric Research, 3080 Center Green Dr, Boulder, USA  
         \and
             Kanzelhöhe Observatory for Solar and Environmental Research, University of Graz, 
             Treffen am Ossiacher See, Austria
         \and
            Instituto de Astrof\'isica de Canarias (IAC), V\'ia L\'actea s/n, E-38205 La Laguna, Tenerife, Spain
         \and
            Departamento de Astrof\'\i sica, Universidad de La Laguna, E-38206 La Laguna, Tenerife, Spain
         \and
            Max-Planck-Institut f\"ur Sonnensystemforschung, Justus-von-Liebig-Weg 3, 37077 G\"ottingen, Germany
             }

  \abstract
   {Large aperture ground-based solar telescopes allow the solar atmosphere to be resolved in unprecedented detail. However, ground-based observations are inherently limited due to Earth's turbulent atmosphere, requiring image correction techniques.}
   {Recent post-image reconstruction techniques are based on using information from bursts of short-exposure images. Shortcomings of such approaches are the limited success, in case of stronger atmospheric seeing conditions, and computational demand. Real-time post-image reconstruction is of high importance to enabling automatic processing pipelines and accelerating scientific research. In an attempt to overcome these limitations, we provide a deep learning approach to reconstruct an original image burst into a single high-resolution high-quality image in real time.}
   {We present a novel deep learning tool for image burst reconstruction based on image stacking methods. Here, an image burst of 100 short-exposure observations is reconstructed to obtain a single high-resolution image. Our approach builds on unpaired image-to-image translation. 
   We trained our neural network with seeing degraded image bursts and used speckle reconstructed observations as a reference. With the unpaired image translation, we aim to achieve a better generalization and increased robustness in case of increased image degradations.}
   {We demonstrate that our deep learning model has the ability to effectively reconstruct an image burst in real time with an average of 0.5 s of processing time while providing similar results to standard reconstruction methods. We evaluated the results on an independent test set consisting of high- and low-quality speckle reconstructions. Our method shows an improved robustness in terms of perceptual quality, especially when speckle reconstruction methods show artifacts. An evaluation with a varying number of images per burst demonstrates that our method makes efficient use of the combined image information and achieves the best reconstructions when provided with the full-image burst.}
   {}


   \maketitle
%

\section{Introduction}
Ground-based solar observations play a crucial role in understanding the magnetic connectivity of the lower solar atmosphere, as they provide unprecedented detail of small-scale structures \citep{EST2022}. However, significant limitations in resolution arise due to the turbulent atmosphere on Earth, lowering the image quality on account of various seeing effects.
State-of-the-art high-resolution solar telescopes are equipped with adaptive optic (AO) systems in order to mitigate seeing effects and reach the diffraction limit \citep{Rimmele2020, Berkefeld2012}. By continuously adjusting a telescope's optics in real time to compensate for atmospheric fluctuations, AO systems significantly improve the resolution of an observation \citep{Rimmele2011}. However, even after AO correction, residual phase errors remain, requiring the application of subsequent image correction techniques to further refine the quality of the observation \citep{Loehfdahl2007}.

The two most used reconstruction methods in the field of ground-based solar imaging are multiframe blind deconvolution, with its extension to multi-object multiframe blind deconvolution, and speckle reconstruction \citep{Loehfdahl2002, Noort2005, Lohmann:83, Woeger2008, Labeyrie1970}. These reconstruction techniques are based on a similar principle. They use the information of a burst of short-exposure images that "freeze" the turbulent atmosphere in each of the frames in order to reconstruct a single high-resolution image \citep{Loehfdahl2007}. The exposure times of these frames must be on the order of milliseconds, depending on the seeing conditions, in order to "freeze" the turbulent atmosphere \citep{Labeyrie1970}. Since the Sun is constantly evolving, the total observation time of a single image burst must not increase the characteristic timescales of the solar features under study to ensure that each frame observes the same state. The typical duration of such image bursts collecting 500 frames is around 20 seconds.
The short-exposure image data obtained by the telescope is given by the convolution of the real object with the point spread function from the optical system and the atmosphere \citep{Loehfdahl2002} as

\begin{equation}
    \label{eq:Image}
    I_{n}(x) = O(x) \ast P_{n}(x).
\end{equation}
Here, $n$ refers to the number of the frame, $I_{n}(x)$ to the raw image burst, $O(x)$ to the real object, and $P_{n}(x)$ to the point spread function from the optical system and the atmosphere.
This is an ill-posed problem since neither the point spread function nor the real object is known. Consequently, the reconstruction process, including finding the point spread function, makes currently used state-of-the-art techniques computationally expensive and time-consuming. In addition, under bad seeing conditions, the reconstruction error increases, which may result in reconstruction artifacts or even failure of the method.

To accelerate the reconstruction process, recent studies have explored different neural network techniques. \cite{Asensio2018} applied a deep convolutional neural network to reconstruct high-resolution ground-based solar observations. The authors used an encoder-decoder and recurrent architectures. \cite{Asensio2023} introduced a more physics-based image reconstruction approach focusing on the determination of wavefront coefficients to construct an accurate wavefront for the deconvolution process. These methods demonstrated the ability to perform state-of-the-art reconstructions in near real time. A shortcoming of these methods is that the reconstruction was applied to a small field of view (FOV) and not to the full FOV observations provided by the telescope.

\begin{figure}[h!tbp]
    \centering
    \includegraphics[scale=0.28]{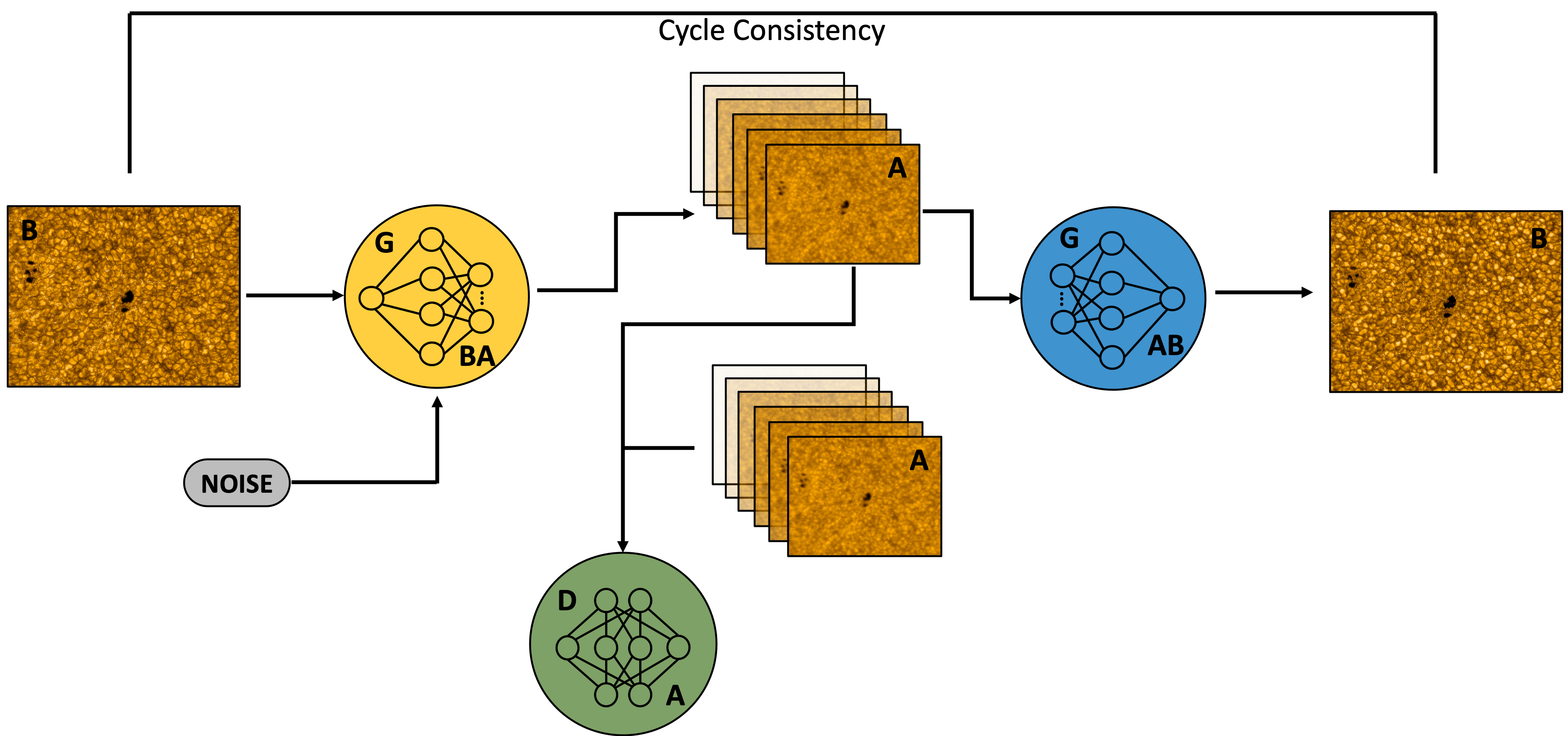}
        \caption{Training cycle for the generation of low-quality image bursts. High-quality images from domain B are translated into a low-quality burst consisting of 100 images by generator BA (yellow). Discriminator A (green) enforces a mapping in the low-quality domain by distinguishing between synthetic (top) and real image bursts (bottom). To fulfill the cycle consistency criteria, generator AB (blue) translates the synthetic image burst back into a single high-quality image of domain B.}
\label{Fig:BAB}%
\end{figure}
Depending on the quality of the bursts, computer memory, and reconstruction time, not all observed image bursts are reconstructed using post-processing methods. This is a major limitation when building a large and diverse dataset for neural network training. Similarly, post-processing methods can lead to artifacts when applied to bursts of reduced image quality. This directly limits paired image translation (e.g.,\citealp{Isola2016}) methods to good seeing conditions. Unpaired image translations enable translations of image domains without additional spatial or temporal alignment while achieving results that are comparable to paired image translation approaches \citep{Zhu2017}. This also increases the training dataset and therefore the applicability to use cases with limited data availability. In case of increased atmospheric seeing conditions, image bursts are often not reconstructed, which leads to an underrepresentation of training samples with low quality.

In the framework of this study, we build on the instrument-to-instrument translation (ITI) method developed by \cite{Jarolim2024}. This method provides a general framework for unpaired image-to-image translation based on generative adversarial networks (GANs; \citealp{Goodfellow2014}). It has been successfully demonstrated to translate between space-based and ground-based solar imaging instruments for image enhancement, instrument intercalibration, and super-resolution observations.
Our goal is to transform images from a low-quality domain (domain A) to a target high-quality domain (domain B). The high-quality domain refers to the speckle reconstruction and serves as a reference for our reconstruction, while the low-quality domain (the level 1 image bursts) serves as the low-quality input. Therefore, we translate 100 short-exposure images into a single high-quality image. The primary training cycle (B-A-B) is shown in Fig.\,\ref{Fig:BAB}.
This training approach encompasses two types of mappings: one from domain A to domain B (A-B) and a second from domain B to domain A (B-A). In a second cycle (A-B-A), we make use of an additional neural network to estimate the noise (noise estimator) of the low-quality image burst. With this we aim to separate the noise of high-quality images, resulting in a performance increase. A third cycle corresponds to an identity mapping, translating images from domain A to domain A and images from domain B to domain B. The training cycles are the same as described in the original ITI paper \citep{Jarolim2024}.

In this study, we apply our method to solar observations from the 1.5 m GREGOR telescope \citep{Schmidt2012_GREGOR, Kleint2020}. A detailed description of the preprocessing steps that were performed is given in Sect.\,\ref{sec:data}. In Sect.\,\ref{sec:method}, we describe the network architecture, the model training procedure, the loss functions employed, and the specific training parameters used. Section \ref{sec:results} presents the results of our model and an evaluation of its performance. The results are further discussed in Sect.\,\ref{sec:discussion} and \ref{sec:conclusion}.

\section{Data}
\label{sec:data}

In this study, we use data from the $1.5$\,m aperture GREGOR telescope located on the Canary Island of Tenerife. It is equipped with the GREGOR Infrared Spectrograph (GRIS; \citealp{Collados2012}), the Broad-Band Imager (BBI; \citealp{Luehe2012}), and the High-Resolution Fast Imager (HiFI; \citealp{Kuckein2017, Denker2023}).

We made use of the HiFI instrument, which encompasses six wavelength bands in total, including Ca\RomanNumeralCaps{2} H, H$\alpha$\,(broadband), H$\alpha$\,(narrowband), G-band, Blue continuum, and TiO. The instrument captures images with a FOV measuring $64.8''\,\times \,54.6''$ and a high-resolution image size of $2560\,\times \,2160$ pixels for the Blue continuum and the G-band channel. Thus, the spatial sampling corresponds to $\sim0.02''$\,pixel$^{-1}$, but it has changed slightly over the years due to re-arrangements of the optical setup. Since the G-band and Blue continuum have been observed the longest and therefore have the largest amount of data, we chose to apply ITI to these two wavelength bands.

\begin{table}[h!tbp]
\caption{Training dataset with level 1 image bursts and level 2 speckle reconstructions.}             
\label{table:dataset}      
\centering                          
\begin{tabular}{c | c | c }        
\hline                 
 & \# of Level 1 bursts & \# of Level 2 speckle \\    
\hline\hline     
G-band & 1078 & 745  \\      
\hline
Blue continuum & 960 & 392 \\
\hline                                   
\end{tabular}
\end{table}

The data is accessible through the Leibniz Institute of Astrophysics (AIP), registration is needed at their website.\footnote{https://gregor.aip.de/}The archive serves as a repository for the original image bursts (level 1), which are made accessible after a one-year embargo. Level 1 data are processed image bursts that have undergone dark- and flat-field corrections using the data reduction package sTools by \cite{Kuckein2017} and are subsequently stored as FITS files within the data archive. Each observation consists of 500 short-exposure frames per wavelength, but only the best 100 are stored. To produce Level 2 data, the 100 best selected images are restored using the speckle masking reconstruction technique \citep{Woeger2008}. 

\begin{table}[h!tbp]
\caption{Test set of failed speckle reconstruction.}             
\label{table:failed_dataset}      
\centering                          
\begin{tabular}{c | c }        
\hline                 
 & \# of failed Level 2 speckle \\    
\hline\hline     
G-band & 170  \\      
\hline
Blue continuum & 107 \\
\hline                                   
\end{tabular}
\end{table}

\begin{table*}[h!tbp]
\caption{Quality metric comparison between the baseline and our ITI method.}             
\label{table:qualmetric}      
\centering                          
\begin{tabular}{c | c c c | c c c | c c c}        
\hline\hline                 
 & & PSNR & & & SSIM & & & MAE & \\    
100 Frames & Baseline & & ITI & Baseline & & ITI & Baseline & & ITI \\       
\hline
& & & & & & & & & \\
G-band & 16.23\,$\pm$\,1.78 & & 20.30\,$\pm\,1.61$ & 0.36\,$\pm$\,0.04 & & 0.66\,$\pm\,0.07$ & 0.13\,$\pm$\,0.02 & & 0.07\,$\pm\,0.01$ \\
& & & & & & & & & \\
Blue Continuum & 14.4\,$\pm$\,2.05 & & 18.20\,$\pm\,1.90$ & 0.33\,$\pm$\,0.07 & & 0.65\,$\pm\,0.07$ & 0.16\,$\pm$\,0.04 & & 0.09\,$\pm\,0.02$ \\
\hline
 & & & & & & & & & \\    
10 Frames & & & & & & & & & \\       
\hline              
& & & & & & & & & \\
G-band & 16.23\,$\pm$\,1.78 & & 18.20\,$\pm\,1.87$ & 0.36\,$\pm$\,0.04 & & 0.55\,$\pm\,0.04$ & 0.13\,$\pm$\,0.02 & & 0.09\,$\pm\,0.02$ \\
& & & & & & & & & \\
Blue Continuum & 14.4\,$\pm$\,2.05 & & 17.19\,$\pm\,1.68$ & 0.33\,$\pm$\,0.07 & & 0.52\,$\pm\,0.08$ & 0.16\,$\pm$\,0.04 & & 0.11\,$\pm\,0.02$ \\

\hline    
\end{tabular}
\tablefoot{Comparison of the full-image bursts (100 frames) and ten frames as input using the test set. ITI shows an improvement in all metrics for both wavelengths, G-band and Blue continuum. The error range corresponds to $\pm\,1\,\sigma$.}
\end{table*}

The  training dataset for both channels consists of observations over a period from 2016 to 2022 (see Tab.\,\ref{table:dataset}). The process of selecting level 1 bursts involved first collecting all available data in the channel, and this was followed by a visual inspection to identify and exclude observations affected by instrumental errors. We followed the same procedure for the reference speckle reconstruction. We note that the dataset size is further reduced due to the additional exclusion of errors caused by the speckle reconstruction method and due to bad seeing observations.

\begin{table}[h!tbp]
\caption{Comparison of the FID quality metric between the baseline and our ITI model.}             
\label{table:FIDqualmetric}      
\centering                          
\begin{tabular}{c | c c c}        
\hline\hline                 
& & FID &  \\    
& Baseline & & ITI \\       
\hline
& & &  \\
G-band & 37.69 & & 6.99 \\
& & &  \\
Blue Continuum &  49.72 & & 5.22 \\

\hline                                   
\end{tabular}
\tablefoot{The comparison is made using the test set. ITI shows an improvement in the FID for both wavelengths, G-band and Blue continuum.}
\end{table}

We built a separate test set consisting of 100 level 1 bursts and speckle reconstructions independent of the training set. We therefore selected the remaining observations with the same requirements as for the training set. For the two channels, this covers four observation days. The test set is independent of the training and the validation set. These dates were selected with a minimum separation of 25 days before and after the training and validation set. In the G-band, the training and validation dataset consists of 1078 level 1 image bursts (29 observation days) and 745 speckle reconstructions (11 observation days). In the case of the Blue continuum channel, we have slightly less data available. Here, the entire training and validation dataset is given by 960 level 1 image bursts (30 observation days) and 392 speckle reconstructions (five observation days).

\begin{figure}[h!tbp]
   \centering
   \includegraphics[scale=0.05]{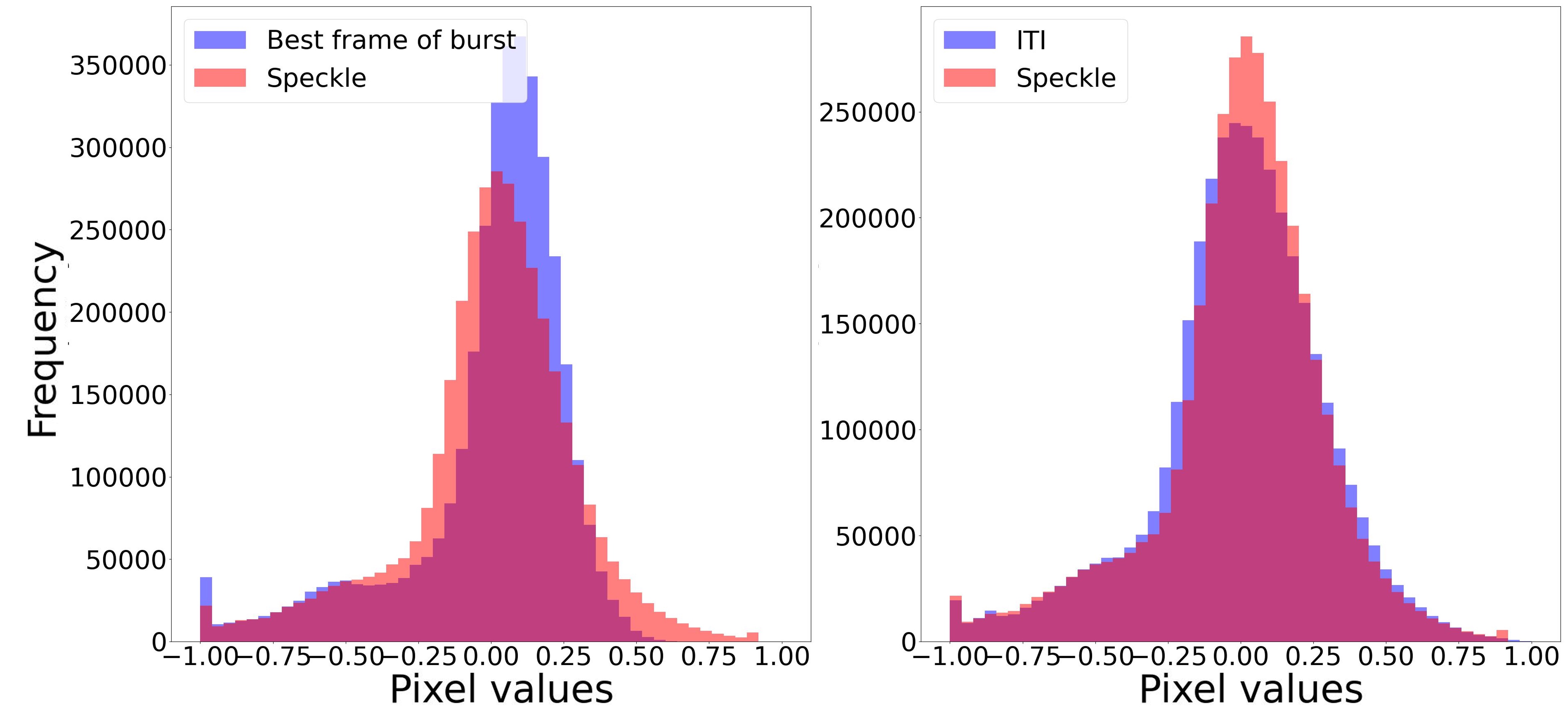}
   \caption{Comparison of the image distribution between a single frame of the original burst and the speckle reconstruction (left) and our ITI reconstruction with the speckle reconstruction (right) from an observation on 11 April 2016.}
    \label{fig:image_distribution}
\end{figure} 

To evaluate the model's performance on observations with lower-quality speckle reconstructions, we created a separate test set containing the original burst with its associated speckle reconstructions. This test set consists of 170 speckle reconstructions in the G-band and 107 speckle reconstructions in the Blue continuum (see Tab.\,\ref{table:failed_dataset}). It is comprised of observations where the speckle reconstruction shows reduced quality, including artifacts.

For the training of the neural network, we used the level 1 image bursts as low-quality input and the level 2 speckle reconstructions as the high-quality reference. 
Due to the fact that our architecture is fully convolutional, we could perform the training with image patches. The evaluation was performed with the full-resolution images. Training a neural network with high-resolution image data can exceed the computational limit, depending on the graphics processing unit (GPU) and memory available. Through the selection of random patches, we extracted diverse patches with 512\,$\times$\,512 pixels from the high-resolution images. This approach offers a twofold benefit, as it facilitates the training process and simultaneously expands our dataset, enhancing the overall quality and robustness of our model. 

In order to have a uniform input for the level 1 image bursts as well as for the level 2 speckle reconstructions, we applied a contrast normalization on the full-resolution images to avoid different levels of brightness and contrast. The spatial sampling of the GREGOR instrument has changed over the years. To have a consistent input, we fixed the spatial sampling for all observations to $0.0276$\, arcsec pixel$^{-1}$ for both the level 1 and level 2 data. For the training of the neural network, we cropped the speckle reconstructions from $2560\, \times \,2160$ pixels to $1800\,\times \, 1500$ pixels. This cropping ensures that only the highest-quality images are used in the training process. This is because the AO system targets specific areas, such as pores or significant regions, resulting in higher image noise at the boundary compared to the locked-on region \citep{Rimmele2011}. The last step was to normalize the input data to the interval of $[-1, 1]$. We normalized the original short-exposure bursts according to the minimum and maximum value. For the speckle reconstructions, we used quantile normalization between 0.001 of the minimum value and 0.999 of the maximum value. This approach appeared to provide better contrast than a minimum-maximum normalization for training. This step is crucial, as it ensures that the input data is in line with our output activation function \citep{Huang2020}.

\section{Method}
\label{sec:method}

Generative adversarial networks have revolutionized the field of computer vision and continue to drive groundbreaking advancements in generating realistic images and improving the quality of existing ones \citep{Goodfellow2014}. GANs have been applied for different applications, such as for super-resolution \citep{Ledig2016} and image synthesis and manipulation \citep{Huang2017}. In solar physics, GANs were applied for solar image restoration by \cite{Jia2019} as well as for solar image generation (e.g., \citealp{Kim2019}, \citealp{Park2019},  \citealp{Shin2020}, \citealp{Jeong2020}, \citealp{Son2021})

A classical GAN architecture consists of two neural networks, a generator and a discriminator. The generator takes as input a random vector from a prior distribution (latent space) and maps it to a target domain. The discriminator takes as input a real sample from the intended domain and a synthetic sample from the generator. The role of the discriminator is to distinguish between synthetic samples from the generator and real samples from the domain. The training is performed in a competitive setup between the generator and discriminator networks. Optimizing the generator involves producing realistic samples within the domain such that the discriminator can no longer distinguish between real and synthetic images. Optimizing the generator to produce realistic samples as estimated by the discriminator completes the competitive training setup. To enable image-to-image translation, the random input vector can be effectively substituted with an image, facilitating translations between images \citep{Isola2016}. Compared to paired image translation, where input and output pairs are required, unpaired image-to-image translation is independent of a spatial and/or temporal overlap of the two datasets. The goal is to relate two data domains, A and B, in order to allow one to translate between two different image domains \citep{Zhu2017}. 

\begin{figure}[h!tbp]
    \centering
    \includegraphics[scale=0.049]{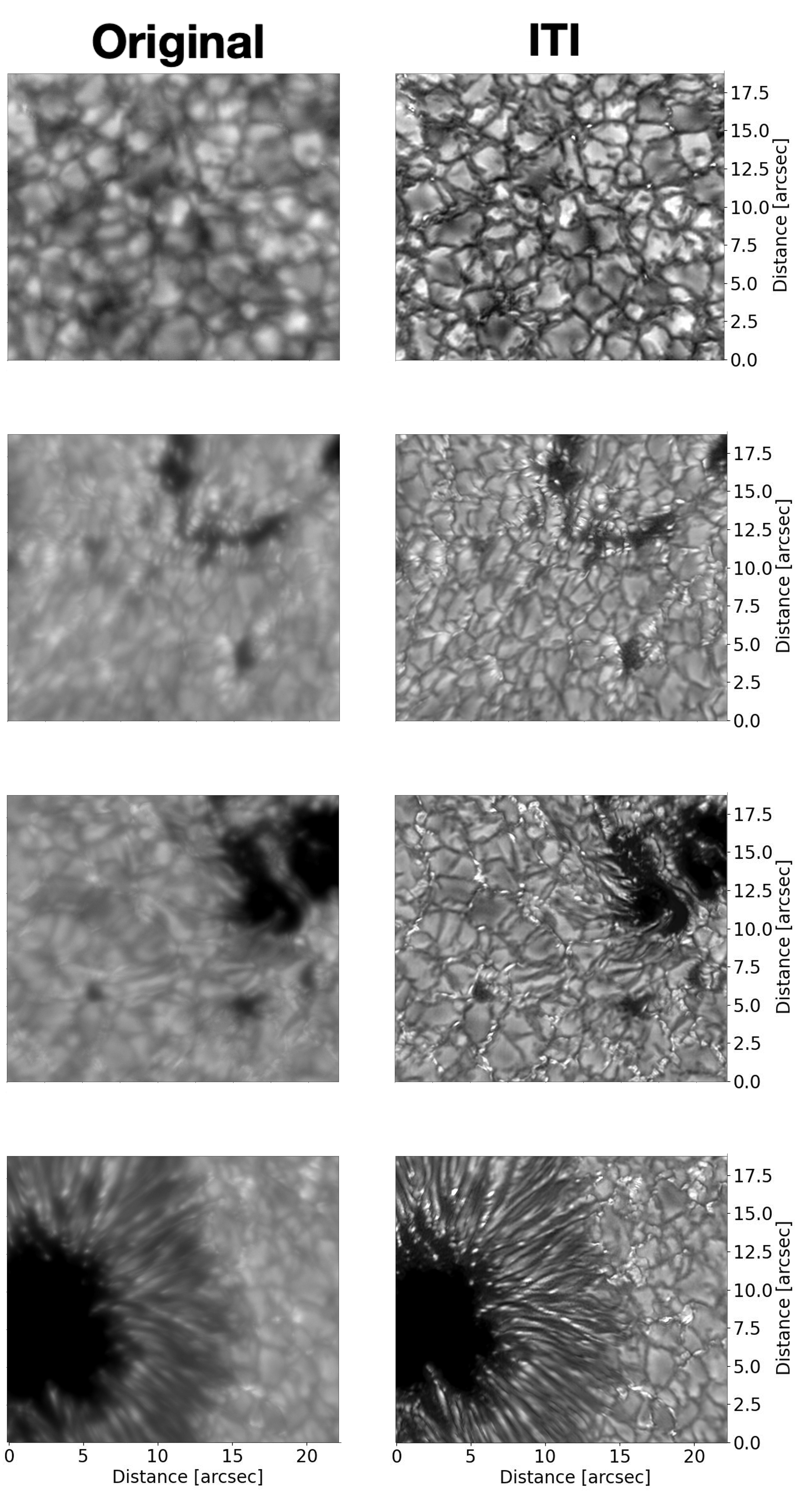}
    \caption{Results of our ITI reconstructions for different solar features. Left: Single frame of original burst in the G-Band ($430.7$\,nm) channel. Right: Corresponding ITI reconstructions.}
    \label{fig:gband_collection}
\end{figure}c

We used the ITI method, which is based on an unpaired image-to-image translation approach \citep{Jarolim2024}. The method translates low-quality images (domain A) to high-quality images (domain B), and it involves synthesizing realistic low-quality images from high-quality images as well as the translation of low-quality images to the target high-quality images.
The model setup includes four neural networks: two generators and two discriminators. The model architectures as well as the training cycles are the same as in the study by \cite{Jarolim2024} except for the number of input channels. The generators have a U-Net architecture that performs downsampling and upsampling operations on the images with skip connections. The discriminators are multi-scale discriminators, as introduced by \cite{Wang2017}. They consist of three individual networks that operate on different scales. The generators learn mappings between domains A and B (A-B and B-A), while the discriminators distinguish between synthetic and real images in both domains. The training process involves taking high-quality images (B), degrading them to domain A using generator BA, and then restoring them back to domain B with generator AB for the first training cycle (see Fig.\,\ref{Fig:BAB}). The second training cycle involves the same process but in reverse order. First, the original image burst is translated with generator AB to a high-quality high-resolution observation. Discriminator B distinguishes between synthesized and real speckle reconstructions. To complete the training cycle, the synthesized speckle reconstruction is translated back to the original image burst with generator BA and the noise estimated from the noise estimator (NE) network. The third training cycle corresponds to an identity mapping where generator AB is used to translate images from domain B to domain B and generator BA is used for the translation of an image burst of domain A to an image burst of domain A.
The total loss is composed of the reconstruction loss, which is used for the distortion quality optimization, and the adversarial loss, for the perceptual quality optimization. The reconstruction loss is subdivided into three components: a mean absolute error (MAE) loss and a content loss, which is an MEA loss but in feature space, evaluating the content of the images. Additionally, it contains a noise loss that is estimated between the sampled noise and the noise estimator. The adversarial loss is computed as proposed by \cite{Mao2016} (least-squares GAN). All individual loss terms are weighted by lambda parameters, which has such advantages as adjusting sensitivity or task specific tuning. 

\begin{figure}[h!tbp]
   \centering
   \includegraphics[scale=0.09]{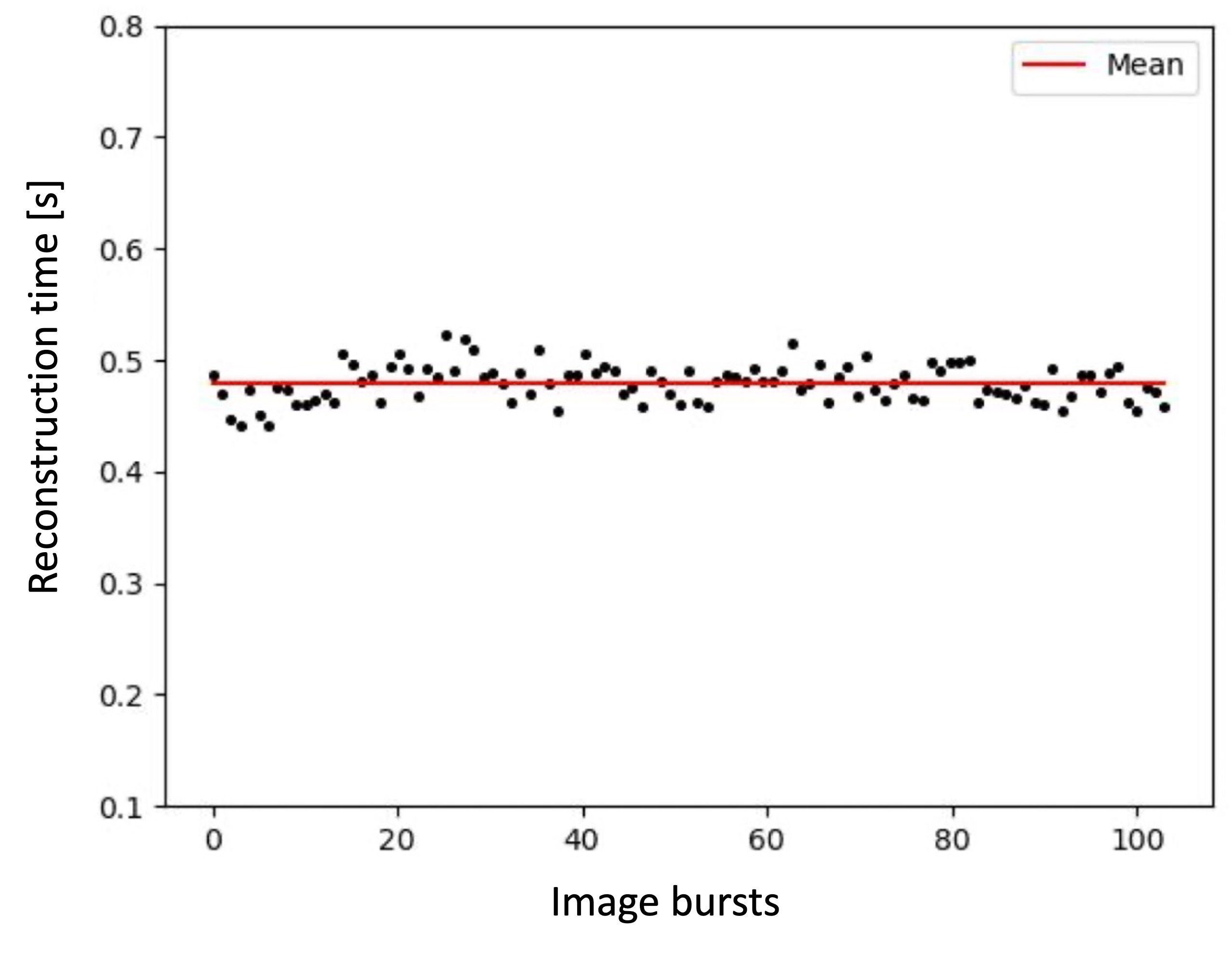}
   \caption{Reconstruction time for image bursts containing 100 frames using a NVIDIA A100 GPU. The red line marks the average over the individual reconstructions.}
\label{fig:translation_time}%
\end{figure}

\begin{figure}[h!tbp]
   \centering
   \includegraphics[scale=0.15]{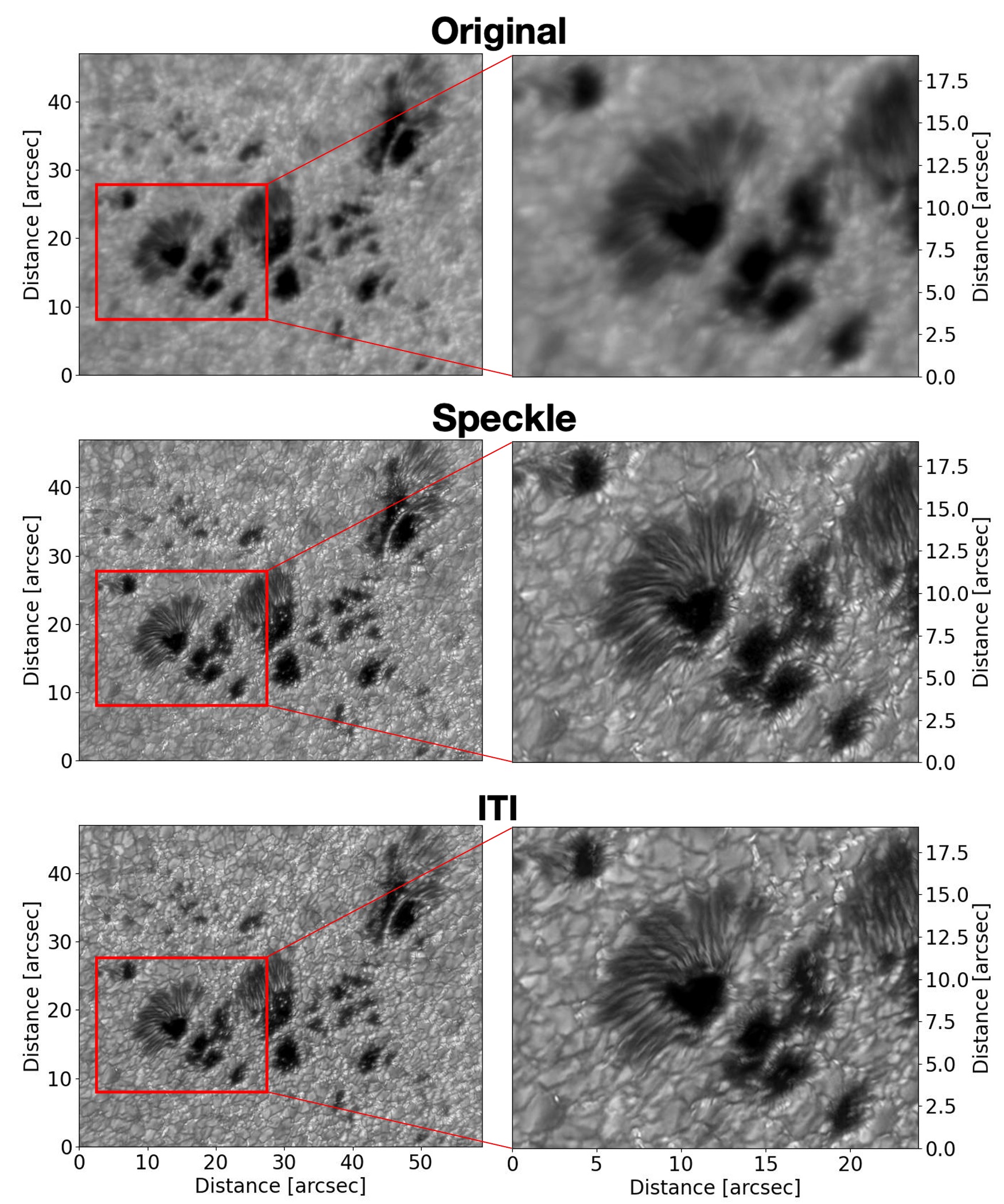}
   \caption{Comparison between a single frame of an original image burst, the speckle reconstruction, and our ITI reconstruction from a G-band observation on 11 April 2016. In the first row, a single frame of the burst is shown with a zoomed-in view marked with a red rectangle. The second row shows the corresponding speckle reconstruction. In the third row, we visualize the restoration of our ITI method, again with a zoomed-in view covering the same spatial region.}
\label{fig:gband}%
\end{figure}

Regarding the discriminator network used by \cite{Jarolim2024}, we adapted it for our application. Since we used a burst of 100 images as input, we added a random multi-scale discriminator. It randomly selects one image from the burst of 100 images to be used for the comparison between synthetic and real observations. In addition, we randomly shuffled the original 100 images before feeding them into the generator network AB. This prevented the generator network from learning a pattern from the image burst. 

\begin{figure*}[h!tbp]
   \sidecaption
   \includegraphics[width=12cm]{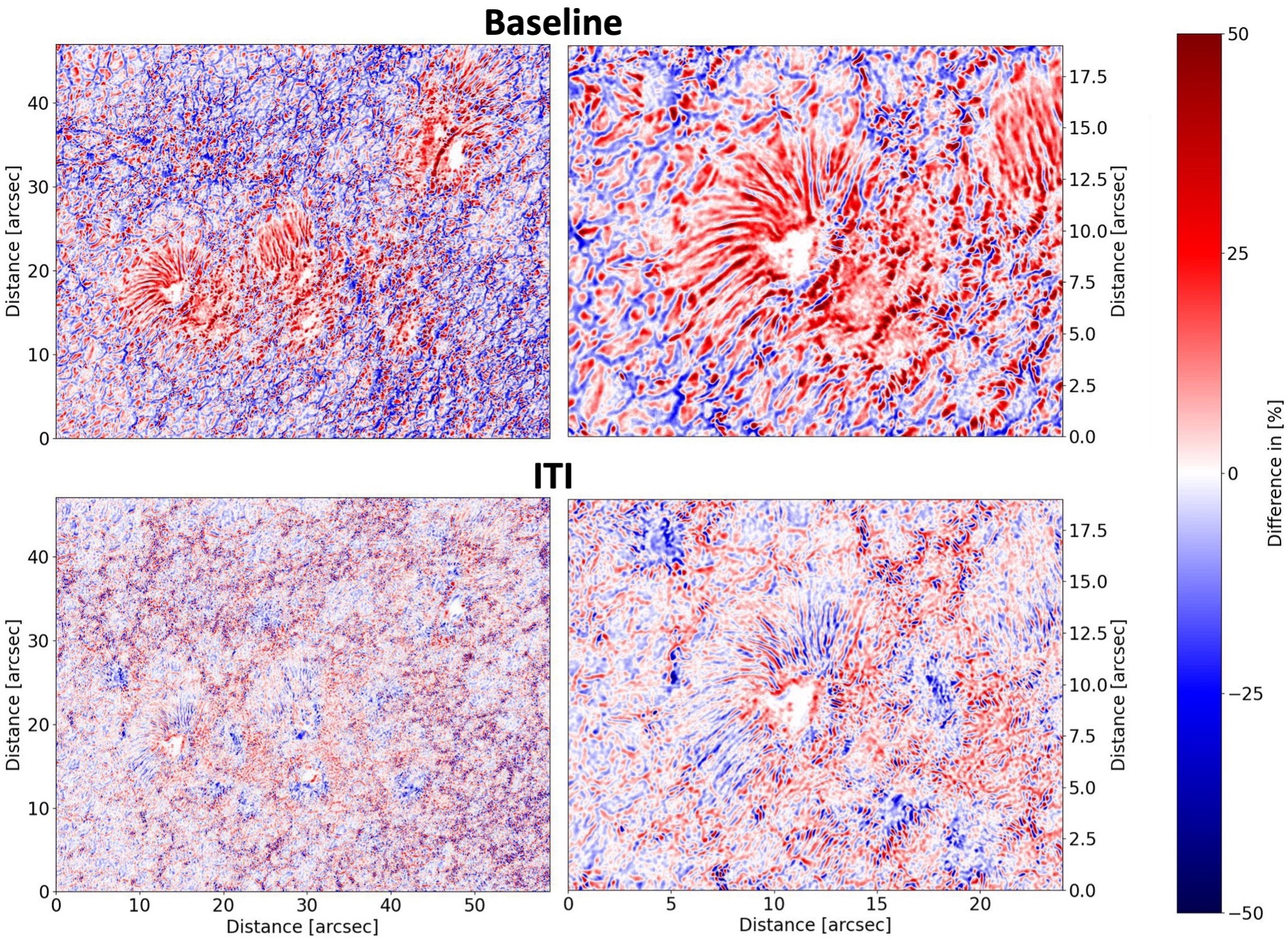}
   \caption{Example to illustrate the performance of an image reconstruction with ITI (lower panels) and the corresponding baseline (difference between single frame of original burst and speckle reconstruction; upper panels). Each panel shows the pixel-wise difference of the reconstructed image with a single frame of the image burst and is called to $\pm 50\,\%$. The right panels show a zoom-in of one of the sunspots shown in the left panels.}
    \label{fig:diff_gband}
\end{figure*}  

\begin{figure}[h!tbp]
   \centering
   \includegraphics[scale=0.15]{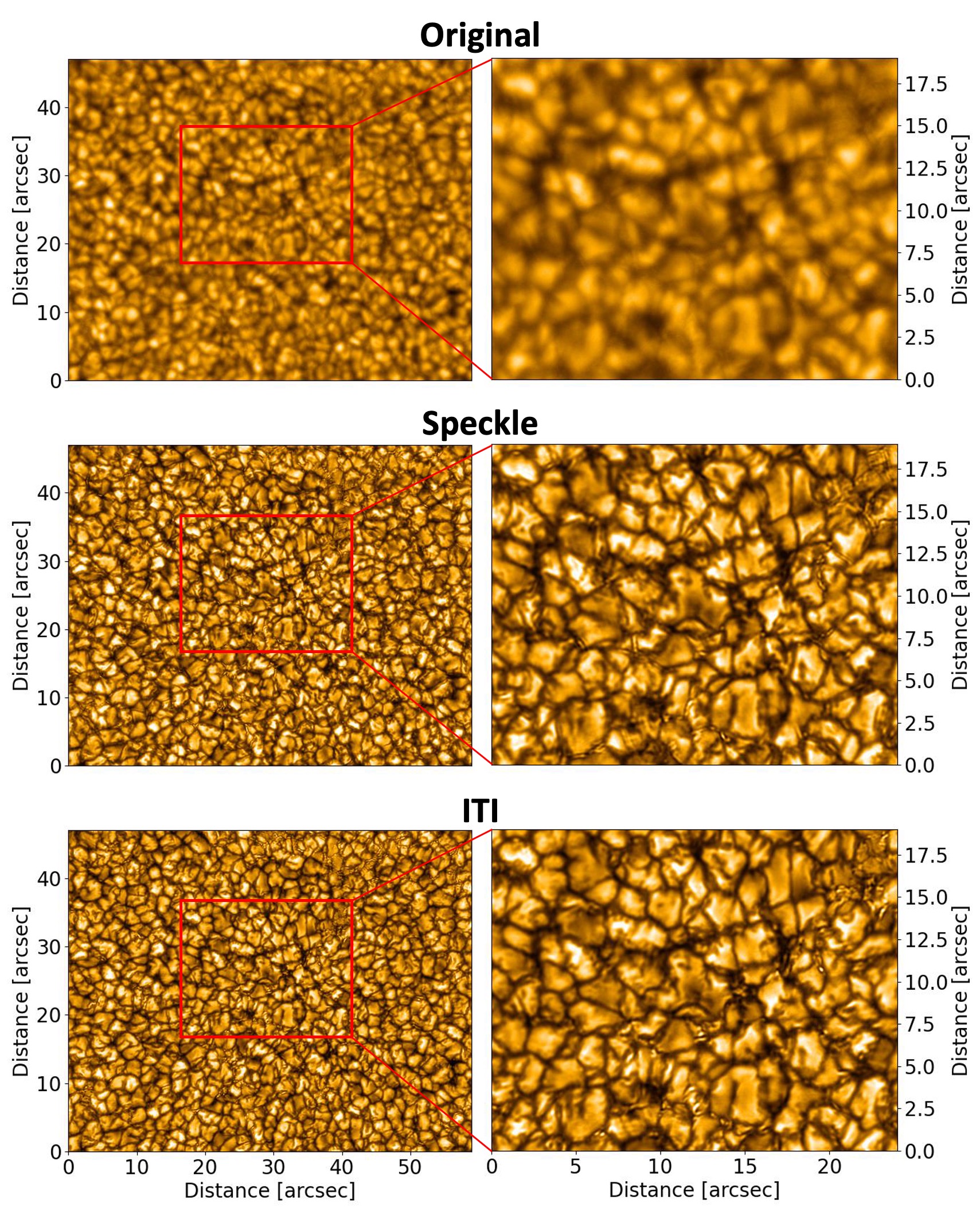}
   \caption{Visualization of a single frame from the original image burst on 13 June 2019 in the first row. The second and third rows show the speckle and ITI reconstruction, respectively. The first column shows the full-resolution observation, and the second column presents a zoomed-in view of the area marked with a red rectangle.}
\label{fig:bcont}
\end{figure}

To optimize the training procedure, we employed the Adam optimizer with a learning rate of 0.0001 \citep{Kingma2014}. The coefficients $\beta_1$ and $\beta_2$ utilized for computing the running averages of the gradient and its square were set to 0.5 and 0.9, respectively. For the instance normalization, we maintained running estimates of the learnable parameters ($\gamma$, $\beta$) for the initial 100,000 iterations, and we fixed them for the remaining iterations. The training was performed with a batch size of one and on image patches with a resolution of $512\,\times\,512$ pixels, which were randomly extracted from the full-resolution observations. In total, about four independent patches could be extracted from a single image. We optimized our model for 220,000 iterations, which corresponded to $\sim51$ and $\sim57$ epochs for the G-Band and Blue continuum, respectively. The $\lambda$ parameters were all set to one by default. We set the identity parameters to 0.1 and gave more weight to content loss by increasing the parameter to ten ($\lambda_{content,id}=0.1$, $lambda_{mae,id}=0.1$, $\lambda_{content}=10$).

In order to provide a comprehensive comparison of the performance between the speckle reconstruction and our model, we evaluated the following quality metrics: mean absolute error (MAE), mean squared error (MSE), structural similarity index measure (SSIM), peak signal-to-noise ratio (PSNR), and the Fréchet inception distance (FID).
The PSNR is defined as the ratio of the maximum possible power of a signal to the power of the distorting noise and is often used as a quality measurement for image data \citep{Fardo2016}.
The SSIM Index is derived from the calculation of three components: the luminance term, the contrast term, and the structural term. The composite index is obtained by multiplying these three components. The SSIM evaluates image degradation by detecting changes in structural information between a reconstructed and a reference image. This quality metric ranges from zero to one, with values closer to one indicating higher quality \citep{Wang2004}.

In addition to the pixel-based metrics, we added another perceptual quality metric to our performance comparison. The Fréchet inception distance (FID) was developed specifically for generative models, such as GANs, to evaluate the perceptual quality of synthesized images. This metric is built on the inception score, which evaluates only the distribution of generated images. In opposition to that, the FID compares the distribution of real images with the distribution of generated images \citep{Heusel2017}. This metric can be used to evaluate an image distribution that shares no spatial and/or temporal overlap between two datasets. Consequently, the FID can be used, for example, to compare a dataset showing artifacts with a high-quality dataset.

\section{Results}
\label{sec:results}

In this study, we trained a neural network on speckle reconstructions from the HiFI instrument mounted on the $1.5$\,m GREGOR telescope. The training was performed for two wavelength bands, with the G-band at $430.7$\,nm and the Blue continuum at $450.6$\,nm. Consequently, we derived two distinct models, each corresponding to a wavelength. We note that a combined training of both channels is not possible for the time range from 2016 to 2022 due to the limited amount of observations that cover both bands simultaneously.
After training the neural network, we evaluated our models on a test set that is independent of the training and the validation set. It contains quiet Sun regions, pores, and sunspots. The test set includes both the level 1 image burst and the corresponding level 2 speckle reconstruction, thus allowing for a direct comparison. In contrast to the training process, which was performed with image patches, the evaluation was carried out with the full-resolution images (2560\,$\times$\,2160 pixels).

\subsection{Quantitative results}
To quantitatively show the performance of our neural network, we evaluated pixel-based and perceptual quality metrics (PSNR, SSIM, MAE and FID). Table \ref{table:qualmetric} summarizes the results for the two wavelength channels in three quality metrics with PSNR, SSIM, and MAE. As a baseline, we considered the comparison between a single frame of the 100 best selected short-exposure frames of the image burst (first frame) and the speckle reconstruction. This defines our baseline. To compare the performance of ITI with this baseline, we also calculated the difference between the ITI reconstruction and the speckle reconstruction and then derived the performance metrics. We show the results using the whole-image bursts (100 frames) for the translation as well as using only ten frames as input. All three quality metrics show a higher quality for each of the wavelength channels. The highest quality is achieved using the full burst, but with only ten frames ITI throughout still outperforms the baseline. The values represent the average over the channels with an error range of $\pm\,1\,\sigma$. For PSNR and SSIM, higher values correspond to better reconstructions/quality and for MAE, the lower the value the better the reconstructions. Table\,\ref{table:FIDqualmetric} shows the quantitative evaluation for the FID, which shows the same quality improvement for ITI compared to the baseline using the whole image burst as input. 

In Fig.\,\ref{fig:image_distribution}, we show a comparison of the image distributions for the G-band observation on 11 April 2016 for the baseline by plotting the histogram for the single frame of the original burst against the histogram of the speckle reconstruction (left panel) and ITI by plotting the histogram for the reconstructions of ITI and speckle (right panel). To allow for the comparison, we set the pixel range between $[-1, 1]$. We observed that the overlap of ITI and speckle is better than the overlap of the baseline over the full pixel range. This figure also confirms the results in Tab.\,\ref{table:FIDqualmetric}, where ITI shows lower FID values and therefore higher perceptual quality for both wavelength channels.

By making use of deep learning for high-resolution solar image reconstruction, we were able to outperform state-of-the-art methods in terms of reconstruction time. We show in Fig.\,\ref{fig:translation_time} that ITI is able to provide reconstruction times in real time. On an NVIDIA A100 GPU, the reconstruction time for a single burst is 0.48\,s. On a standard AMD CPU, the average reconstruction time is 3.46\,s.

\begin{figure}[h!tbp]
   \centering
   \includegraphics[scale=0.11]{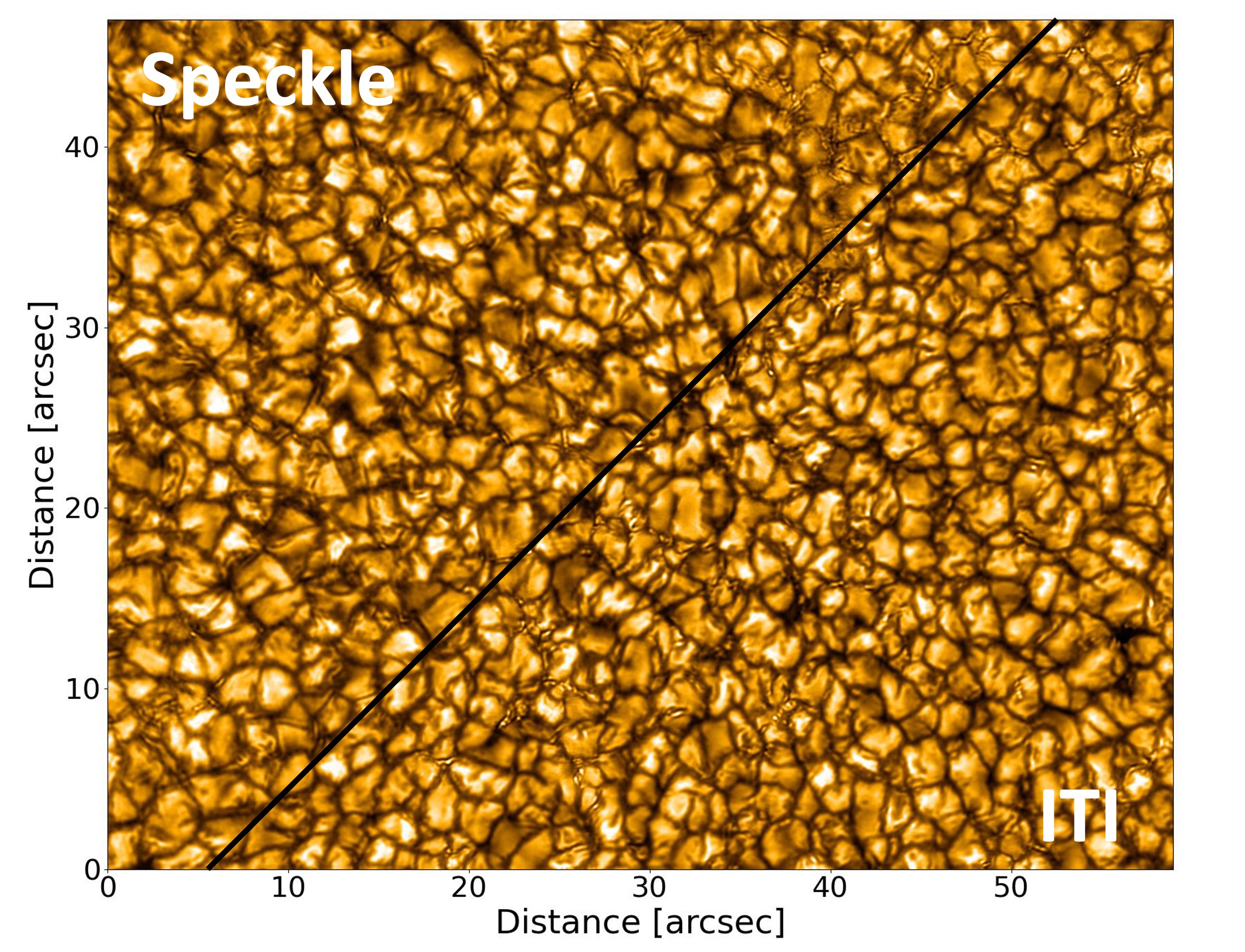}
   \caption{Comparison of the speckle reconstruction with our ITI reconstruction. In the top left, the speckle reconstruction is shown. In the bottom right is the ITI reconstruction.}
    \label{fig:diff_bcont}%
\end{figure}

\subsection{G-band - Qualitative comparison}
\label{sec:gband}
Our first model was trained on the G-band wavelength at $430.7$\,nm. We processed the data as described in Sec.\,\ref{sec:data} and trained the network for 220,000 iterations (51 epochs) until convergence was reached.

Figure\,\ref{fig:gband_collection} shows that ITI can perform reconstructions on a variety of observations. This includes a quiet Sun region (first row), an observation closer to the limb (second row), a pore with a light bridge (third row), and a sunspot (fourth row). For each row, ITI shows detailed structures in the reconstructed observations.
In Fig.\,\ref{fig:gband}, we show an example result for the observation of 11 April 2016. The first row shows a single frame of the original image burst. The second row visualizes the speckle reconstruction, and the third row shows our ITI reconstruction. For a more detailed comparison, zoomed-in views are shown.

The quality difference between the single frame of the burst (first row) and the reconstructions (second and third rows) demonstrates the necessity for post-image reconstruction methods for high-resolution ground-based solar observations. ITI shows a clear improvement and a close resemblance to the reference speckle reconstruction on the full-resolution observation. The zoomed-in view visualizes the performance of our method also on smaller scales. However, we note the ITI reconstruction shows small distortions in the penumbra and enhanced contrast in the pores. This is also confirmed by the enhanced difference in Fig.\,\ref{fig:diff_gband}, where we show a direct comparison between the speckle reconstruction and our ITI reconstruction. The difference map is plotted for the baseline and for ITI in a 50\,\% interval. The left column compares the full-resolution image, and the right column presents the same zoomed-in region as shown in Fig.\,\ref{fig:gband}. A slight difference increase for ITI is also visible for the intergranular lanes, which could be related to the enhancement of small-scale structures close to the resolution limit.

\begin{figure}[h!tbp]
   \centering
   \includegraphics[scale=0.1]{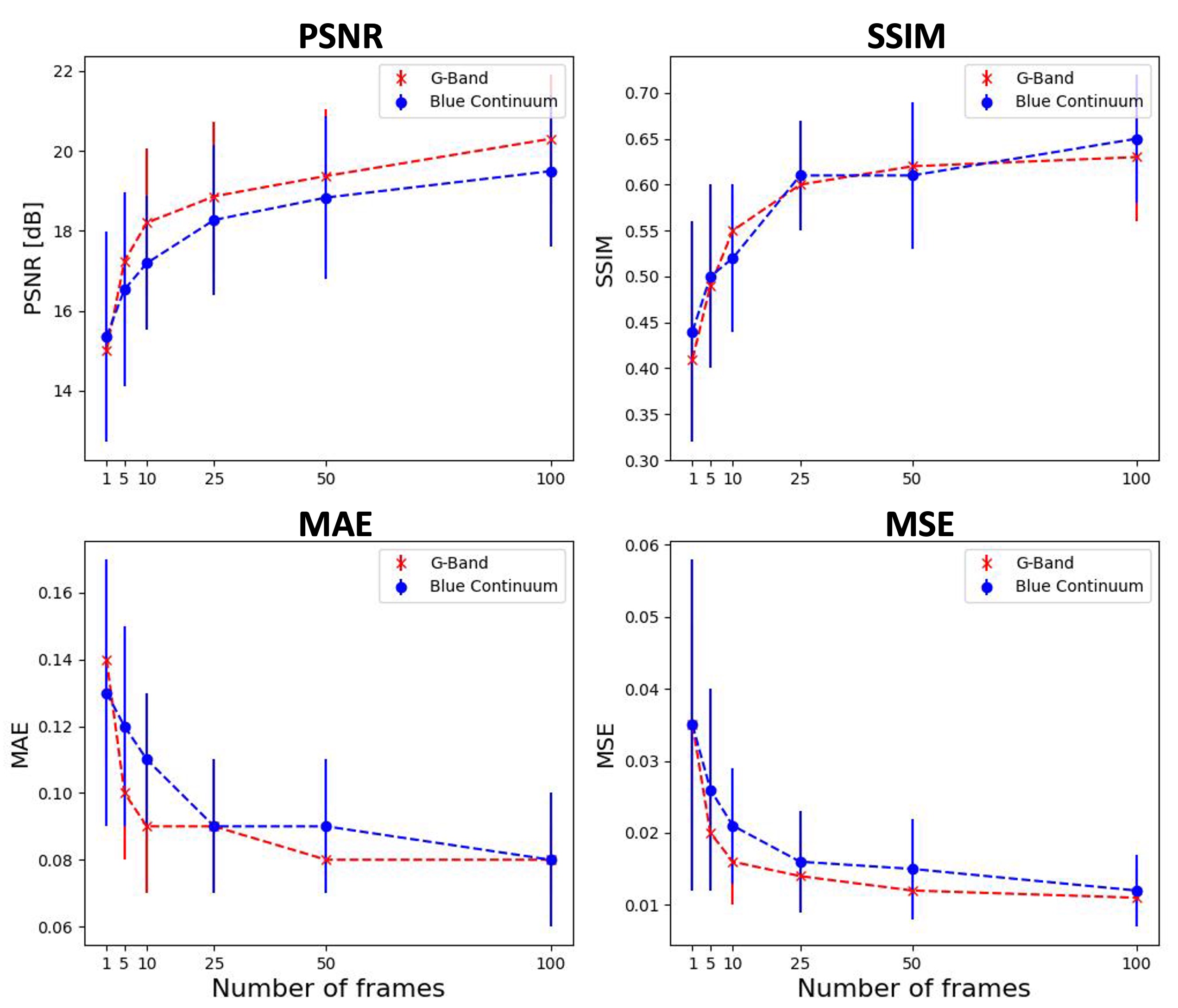}
   \caption{Evaluation of the quality metrics MAE, MSE, SSIM, and PSNR with a varying number of input frames of the burst for both the G-band (red) and Blue continuum (blue). The best quality for all metrics is achieved by using the whole image burst (100 frames) for the reconstruction.}
\label{fig:framecomparison}
\end{figure}

\begin{figure*}[h!tbp]
   \sidecaption
   \includegraphics[width=12cm]{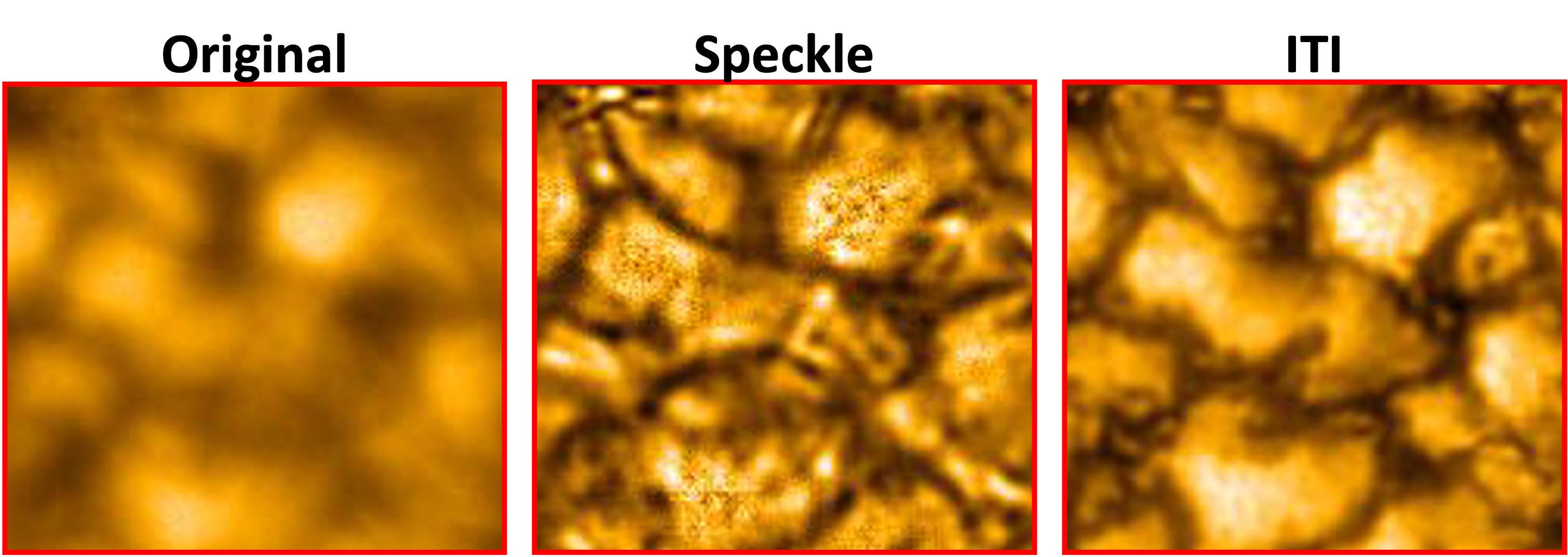}
   \caption{Evaluation of speckle reconstruction showing artifacts in the Blue continuum ($450.6$\,nm) from the 19 October 2019. The first column shows a single frame of the original burst, the second column shows the speckle reconstruction, and the third column shows our ITI reconstruction with a field of $5.5''\,\times \,5.5''$.}
    \label{fig:bcont_artifact}
    \end{figure*}

\subsection{Blue continuum - Qualitative comparison}
Our second model was trained on the Blue continuum spectral band at $450.6$\,nm. We performed our model training and evaluation analogously to the G-band data (Sec.\,\ref{sec:gband}) with 220,000 iterations (57 epochs) and a patch size of 512\,$\times$\,512 pixels. In Fig.\,\ref{fig:bcont}, we show an observation from 13 June 2019. The first row shows a single frame of the original burst. The second row shows the speckle reconstruction, and the third row shows the result of our ITI reconstruction. Again, for each row, the zoomed-in views are shown, marked by the red rectangles, covering the same sized spatial region. The figure demonstrates and confirms the results obtained in the quality metrics comparison shown in Tab.\,\ref{table:qualmetric}. In addition to the image burst-speckle-ITI reconstruction comparison, we plot the speckle reconstruction and our ITI restoration in a single image in Fig.\,\ref{fig:diff_bcont}. Here, the speckle reconstruction and the ITI reconstruction show a strong perceptual similarity, which demonstrates that ITI can achieve a similar image enhancement to the reference speckle images.

\subsection{Burst stacking}
As mentioned above, solving the image equation (Eq.\,\ref{eq:Image}) is an ill-posed problem. We demonstrate that the additional information for the reconstruction is drawn from using the entire image burst, similar to the speckle reconstruction. To investigate whether more input frames improve reconstruction quality, we trained our network with different numbers of input frames. We used the same parameter setting and trained our method with a varying number of input frames. We input sequences of 1, 5, 10, 25, 50, and 100 frames. To evaluate the model performances, we used the high-quality test sets as described in Sec.\,\ref{sec:data}. In all quality metrics with MSE, MAE, PSNR, and SSIM, we observed an improvement in quality with the increase in the number of input frames (Fig.\,\ref{fig:framecomparison}). The data points in Fig.\,\ref{fig:framecomparison} show the average over the test set with an error range of $\pm\,1\,\sigma$ for each sequence and wavelength. For just a single image from the burst as input, the information is not sufficient, and the model produces artifacts. The largest quality increase is achieved between 1 and 25 images as input. The highest quality is achieved by using the whole burst (100 frames) as input. This result demonstrates that our model effectively uses the information of all the 100 short-exposure images of the burst for the reconstruction process. Since originally 500 short-exposure images were taken but only the best 100 are stored, we expected that the quality of the ITI reconstructions would increase even more if a burst of up to 500 short-exposure frames is used as input.
    
\subsection{Failed speckle reconstruction}
We also examined the performance of our model in the presence of artifacts in speckle reconstructions. In addition to the results presented above, this evaluated the robustness and reliability of our model. Speckle artifacts can arise due to various factors, including imperfections in the imaging system, noise, and aberrations, but are mostly due to bad seeing conditions. With the use of unpaired image-to-image translation, we learned the mapping into a high-quality domain that does not include Fourier transformations to obtain the reconstructed image. This reduces the generation of artifacts.

We built an additional test set of level 1 bursts and their corresponding speckle reconstructions containing images with artifacts and not well reconstructed observations as described in Sect.\,\ref{sec:data}. The test set was visually selected and consists of 150 observations in the G-band and 107 observations in the Blue continuum, respectively. We evaluated the reconstruction performance on the high-quality test set (see Sect.\,\ref{sec:data}). We made a comparison between the baseline, the speckle reconstruction showing artifacts, and ITI. Since pixel-based metrics cannot be used for the comparison, as the speckle reconstructions show artifacts, we only used the perceptual quality metric with the FID. 

\begin{table}[h]
\caption{Perceptual quality comparison between the baseline, speckle reconstructions showing artifacts and our ITI reconstructions.}             
\label{table:badspeckle}      
\centering                          
\begin{tabular}{c | c c c c }        
\hline\hline                 
 & &  FID &  \\    
 & Baseline & Low-quality Speckle & ITI  \\       
\hline
& & &  \\
G-band & 43.34 & 17.94 & 13.99 \\
& & &  \\
Blue Continuum & 55.79 & 17.17 & 11.42 \\
\hline                                   
\end{tabular}
\tablefoot{The comparison is made using the FID. For the FID, lower values correspond to a higher perceptual quality.}
\end{table}

The evaluation of the FID in Tab.\,\ref{table:badspeckle} shows that our model is able to reconstruct image bursts with a higher perceptual quality compared to the baseline and the speckle reconstructions containing artifacts. An example is shown in Fig.\,\ref{fig:bcont_artifact} from the observation on 19 October 2019 in Blue continuum. 
We show a single frame of the original burst, the speckle reconstruction, and our ITI reconstruction with a small subfield of $5.5\,''$ x $5.5\,''$. When comparing the original frame with its speckle reconstruction, we observed the appearance of artifacts in the reconstructed image. ITI is able to reconstruct the burst without showing artifacts, or at least they are strongly reduced, leading to a smoother reconstruction and therefore higher perceptual quality. This also confirms the higher perceptual quality in Tab.\,\ref{table:badspeckle}. We note that this increased robustness is facilitated by the unpaired image-to-image translation, where the model learns to translate between image domains rather than individual image pairs. With this unpaired approach we can overcome limitations of the reference speckle method when dealing with challenging reconstructions.

\begin{figure}[h!tbp]
   \centering
   \includegraphics[scale=0.15]{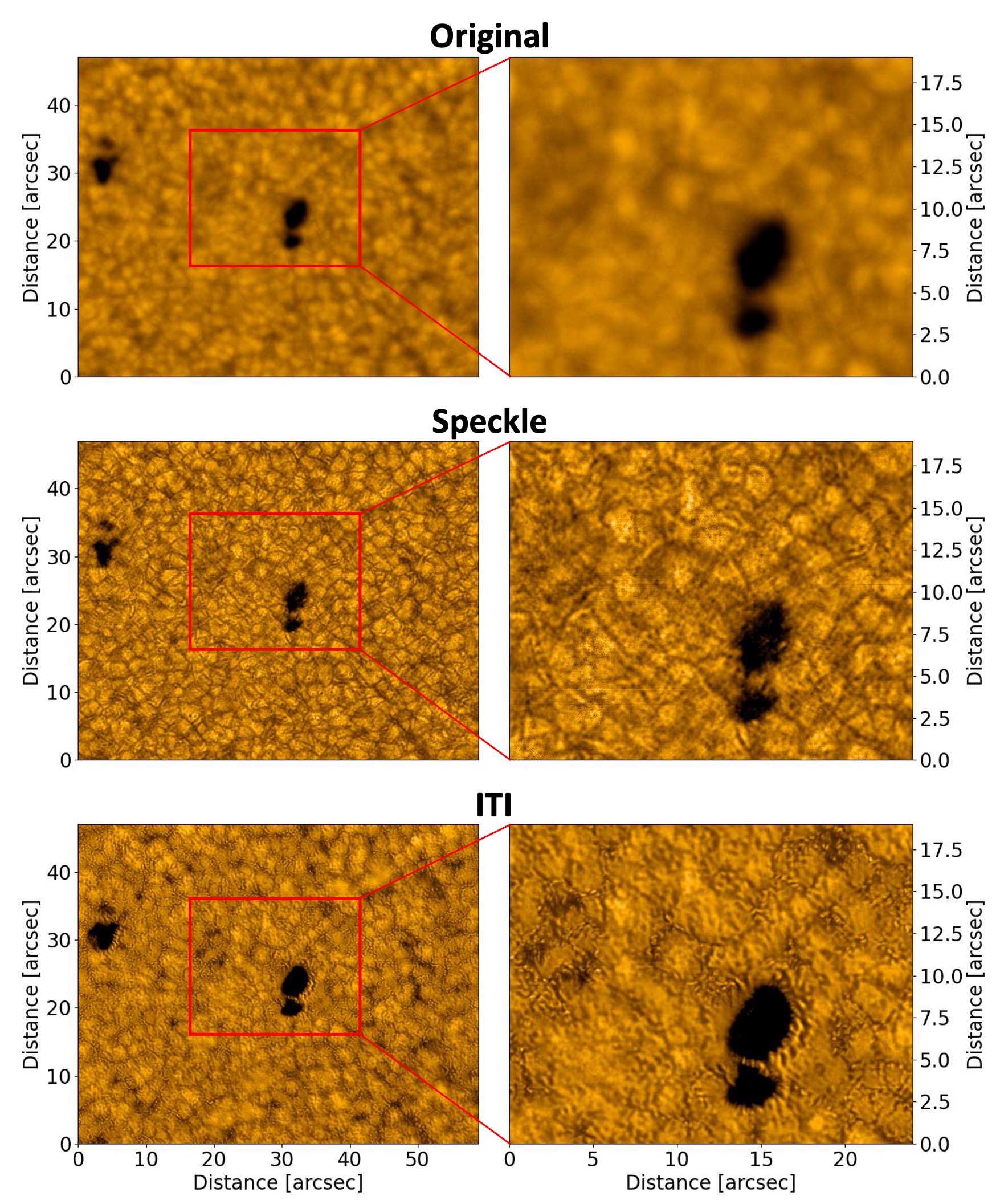}
   \caption{Example where ITI shows limitations in the reconstruction of an image burst with an increased noise level from 29 May 2019. We show a single frame of the original burst in the first row, the speckle reconstruction in the second row, and our ITI reconstruction in the third row, with zoomed-in views.}
              \label{fig:ITI_limit}%
    \end{figure}

\subsection{ITI limitations}

Unlike the classical methods that first evaluate the point spread function and subsequently deconvolve the image burst, our reconstruction method with image domain translation learns a target image distribution for the reconstruction process. The first limitation of this approach is that the image enhancement process is not interpretable. We can only compare the method with state-of-the-art reconstruction techniques. Second, image domains that are outside of our training set cannot be reconstructed. Third, if there are aberrations in the original image burst, ITI is not able to produce sharp reconstructions. An example is shown in Fig.\,\ref{fig:ITI_limit} from an observation on 29 May 2019 in the Blue continuum wavelength band. 
Both the speckle reconstruction and ITI show distortions across the whole image. The ITI reconstruction appears blurred, similar to the single frame of the original burst. The zoomed-in view confirms this observation also on smaller scales.
    
\section{Discussion}
\label{sec:discussion}
We have demonstrated that our ITI approach is able to reconstruct short-exposure image bursts with a quality similar to that of state-of-the-art reconstruction methods and more robustness even for a limited dataset. With the unpaired image-to-image translation approach \citep{Zhu2017, Jarolim2024}, there is no need for an overlap, neither in temporal nor spatial data. Compared to paired image translation, such as \cite{Isola2016}, this approach has a big advantage in overcoming the limitation of data. In many cases, not all original bursts are used for the reconstruction, and therefore not all original bursts can be used for the paired image-to-image translation training. When considering paired image translation to reconstruct a short-exposure burst into a single high-resolution observation, the dataset limitation is given by the number of high-quality speckle reconstructions available \citep[e.g.,][]{Asensio2018}. In our case, this corresponds to 745 speckle reconstructions for the G-band and only 392 for the Blue continuum. Using the unpaired approach, the dataset of input bursts increases to 1078 and 960 observations respectively for the G-band and Blue continuum (increase of $\sim 44\%$ and $\sim 145\%$). In addition, with the unpaired approach, we also used image bursts for training that are not post-processed for various reasons, such as bad seeing conditions. This allowed the network to learn the mapping in the high-quality domain for observations that showed artifacts in the speckle reconstructions, adding more robustness to our method.

When compared to the baseline where we calculate the difference between the single frame of the original burst (first frame) and the speckle reconstruction, ITI can outperform the baseline on all quality metrics (see Tab.\,\ref{table:qualmetric} and \ref{table:FIDqualmetric}), both pixel-based (MAE, MSE) and perceptual (SSIM, FID). As demonstrated, the neural network can produce more robust reconstruction when the speckle data shows artifacts, leading to higher perceptual quality. 

The ITI method makes efficient use of the full-image burst, as can be seen from Fig.\,\ref{fig:framecomparison}, where we compare the model performance as a function of the number of input frames for each reconstruction. The quality increases as the number of input frames increases, reaching the highest quality when the entire burst is translated. However, we note that even when using only ten frames, on average, the metrics are already better than for the baseline. Fig.\,\ref{fig:framecomparison} suggests that more information (a higher number of input frames) improves the quality of the ITI reconstructions. Therefore, applying ITI to more than 100 short-exposure images before saving the best 100 images from the original 500 observations could improve the reconstruction even further. With this we demonstrate that we can mitigate the ill-posed problem and provide a more informed image enhancement. After model training, our method can be applied to full-resolution images (e.g., Fig.\,\ref{fig:gband}). 

A crucial consideration when employing deep learning for the reconstruction of ground-based solar observations is the issue of reconstruction time. Our approach allows us to perform real-time restoration of the original burst, and on average it takes only 0.48 seconds per reconstructed image when utilizing an NVIDIA A100 (see Fig.\,\ref{fig:translation_time}). On a standard AMD CPU, this process takes 3.46 seconds on average. Fast reconstruction times are especially important for large-aperture telescopes, such as the Daniel K. Inouye Solar Telescope (\citealp{Rimmele2020}) and the future European Solar Telescope (\citealp{EST2022}). It is essential for storage purposes as well, as real-time reconstructions can also be used during the observation process itself.

A shortcoming of our method is the limitations in the dataset, which contain only a small number of sunspots. Reconstructions of sunspot observations are therefore less reliable than reconstructions of quiet-Sun regions. Increasing the number of high-quality speckle reconstructions of sunspots in the training set could improve the ITI reconstructions of sunspot umbra and penumbra. While unpaired image-to-image translation with the ITI method allows us to leverage the full data archive independent of paired samples, a larger dataset with an equivalent feature distribution could improve the reconstruction results. For image bursts that show larger aberrations in the original image burst, ITI is not able to produce sharp reconstructions of the burst similar to the speckle reconstruction (see Fig.\,\ref{fig:ITI_limit}).

\section{Conclusions}
\label{sec:conclusion}
We have presented an image reconstruction method for high-resolution ground-based solar observations using unpaired image-to-image translation. The method was applied to the HiFI instrument from the GREGOR telescope and utilizes the information from 100 short-exposure observations for restoration, similar to the speckle reconstruction technique by \cite{Luehe1993}. With reconstruction in real time, this method can be considered as an alternative to the currently used methods, such as speckle reconstruction \cite{Woeger2008} and multi-object multiframe blind deconvolution \cite{Noort2005}.

Despite providing more robust results than state-of-the-art methods, ITI becomes limited when only a small set of image distributions are covered in the training set. The inclusion of recent observations, with their reconstructions of the approaching maximum of solar cycle 25, could further improve the model.
In addition, the neural network architecture requires high-quality reference observations. The current training is based on speckle reconstruction, and therefore the method is constrained by the quality of the reference images. An alternative approach would be to build on the physical equations of image formation, similar to \cite{Asensio2018, Asensio2023}, or to use simulated observations as reference \citep{Rempel2017}.

\section*{Code availability}
The code is publicly available in the following repository: \url{https://github.com/spaceml-org/InstrumentToInstrument}. 

\begin{acknowledgements}
This research has received financial support from the European Union’s Horizon 2020 research and innovation program under grant agreement No. 824135 (SOLARNET) and from the University of Graz EST (European Solar Telescope) program. RJ was supported by the NASA Jack-Eddy Fellowship. CK acknowledges grant RYC2022-037660-I funded by MCIN/AEI/10.13039/501100011033 and by "ESF Investing in your future". We acknowledge the use of the Vienna Scientific Cluster (VSC) for the computational resources and obtaining the results presented in this paper. \\
This research has made use of AstroPy \citep{astropy2022}, SunPy \citep{sunpy2020} and PyTorch \citep{pytorch2017}. 
\end{acknowledgements}

%
%
\bibliography{bib}
\bibliographystyle{aa}
\end{document}